\documentclass[aps,twocolumn]{revtex4}
\usepackage[dvips]{graphicx}
\begin{document}

\title{ The Bogoliubov Theory of a BEC in Particle Representation }

\author{ Jacek Dziarmaga and Krzysztof Sacha}

\affiliation{
Intytut Fizyki Uniwersytetu Jagiello\'nskiego, 
ul. Reymonta 4, 30-059 Krak\'ow, Poland 
}

\begin{abstract}

In the {\it number-conserving} Bogoliubov theory of BEC the Bogoliubov
transformation between quasiparticles and particles is nonlinear. 
We invert this nonlinear transformation and give general expression 
for eigenstates of the Bogoliubov Hamiltonian in particle
representation. The particle representation unveils structure of 
a condensate multiparticle wavefunction. We give several examples
to illustrate the general formalism.

\end{abstract}

\maketitle

PACS 03.75.Fi, 05.30.Jp    

\section{ Introduction }

   Experimental realization of a Bose-Einstein condensate (BEC) in
trapped alkali atoms triggered feverish activity in cold atom physics 
\cite{nobel}. Recent development of experimental techniques in 
trapping and cooling of dilute atomic gases allows for preparation, 
manipulation and control of a condensate in various states and various 
trapping potentials. For instance, excitation of a condensate to vortex and 
solitonic states became possible \cite{vor-sol-exp}. Cold 
atoms are also becoming a convenient toy system to study solid state physics. 
For example, cold atoms trapped in an optical lattice potential undergo quantum 
phase transition to Mott insulator phase when the barrier between minima of 
the lattice potential is increased \cite{Hansch,Kasevich}.

   Theoretical description of a many body system is a challenge to
physicists.  At low temperature our understanding of BEC is usually 
based on a mean field approximation within the Gross-Pitaievskii equation 
(GPE) \cite{GPE}. GPE provides satisfactory experimental 
predictions when depletion of a condensate, induced by interactions
between atoms, is negligible. In order to describe these quantum 
fluctuations around the condensate in a systematic way one can use 
the Bogoliubov theory of BEC. The key idea of the original Bogoliubov 
theory \cite{BT} is the $U(1)$ symmetry breaking approach where 
the atomic field operator is assumed to have a nonzero expectation value. 
This {\it coherent} state necessarily involves superposition of states 
with different numbers of atoms, an assumption very far from experimental
reality.

  A number-conserving Bogoliubov theory has been introduced in 
\cite{gardiner,castin}. While taking into account
a finite number of atoms in a condensate, this theory represents eigenstates
of the Bogoliubov Hamiltonian as Fock states with well defined numbers
of Bogoliubov quasiparticles in all Bogoliubov modes. In this {\it 
quasiparticle} 
representation it is possible, and even pleasant, to calculate a single 
particle density matrix. However, higher order correlators are increasingly 
more painfull to get. For all practical purposes in the quasiparticle
representation the multiparticle correlations remain hidden. The single particle 
density function carries very limited information about a many-body system, 
as is best illustrated by an example considered by Javanainen and Yoo 
\cite{java}. They simulated a measurement of atom density distribution 
in an atomic Fock state $|N,N\rangle$ (it is a model counterpart of an 
experiment with
interference of two separate BECs \cite{interference}). With two 
counter-propagating plane waves as the two modes in the Fock state the 
single particle density function gives uniform density distribution. However, 
sampling of the many-body probability density shows that each realization of the
experiment reveals an interference pattern. Only after averaging over many
realizations one gets the uniform density distribution
predicted from the single particle density function. 

  It seems thus crucial to understand the structure of a condensate 
multiparticle wavefunction especially when the number of non-condensed atoms 
becomes considerable and may play a significant role in an outcome of a 
measurement. In this paper we derive an expression for Bogoliubov eigenstates in 
{\it particle} representation i.e. as superpositions over Fock states with well 
defined numbers of {\it particles} (atoms) in a condensate and non-condensate 
single particle states. For example, the state without any quasiparticles
(Bogoliubov vacuum) has a pair-correlated form
\begin{equation} 
\left(   
\hat a_0^{\dagger} \hat a_0^{\dagger} +
\sum_{m>0}
\lambda_m
\hat a_m^{\dagger} \hat a_m^{\dagger}
\right)^{\frac{N}{2}}~|0\rangle~,
\nonumber
\end{equation}
where the index $m$ runs over a basis of modes $\phi_m(x)$ orthogonal
to the condensate wavefunction $\phi_0(x)$. This pair-correlated ansatz
has been proposed in a review by Leggett \cite{rmp}, but (except for
a homogeneous condensate \cite{rmp}) the eigenvalues $\lambda_m$ and
the eigenmodes $\phi_m$ remain unknown.
In this paper we construct a simple relationship between these 
eigenmodes/eigenvalues and the elementary Bogoliubov excitations of the system 
for a general inhomogeneous condensate which is of current 
experimental interest. The pair correlated vacuum state is a foundation
on which one can build the theory of BEC entirely in terms of $N$-particle
wavefunctions. 

  The paper is organized as follows. In Section~\ref{bogcast} we briefly 
summarize 
the number-conserving Bogoliubov theory introduced by Castin and Dum 
\cite{castin}
for a system with a finite number of particles. In Section~\ref{stanz} we derive 
the particle representation of the Bogoliubov eigenstates. In 
Sections~\ref{2W},\ref{3W},\ref{becharm} we illustrate the particle 
representation
with a series of examples: a double well, a triple well, and a condensate in 
a harmonic potential. We conclude in Section~\ref{concl}.

\section{ Bogoliubov theory with a finite number of particles}
\label{bogcast}

   The original Bogoliubov theory (BT) \cite{BT} was build around the
concept of spontaneous symmetry breaking: the field annihilation operator
$\hat\psi$ acquires a large nonzero expectation value,
$\langle\hat\psi(\vec x)\rangle=\phi_0(\vec x)$. This nonzero mean is
sometimes regarded as a symptom of Bose-Einstein condensation. The
original BT is build by expansion in small quantum fluctuations around
this large mean, $\hat\psi(\vec x)=\phi_0(\vec x)+\delta\hat\psi(\vec x)$.
This method has to be used with a lot of care. For example, careful
application of the BT leads to the divergence of the quantum fluctuations
that is interpreted as a quantum phase spreading of the condensate
\cite{lew96}.  In the original spontaneous symmetry breaking approach the
condensate state is not a stationary state because for
$\langle\hat\psi(\vec x)\rangle \ne 0$ the number of particles is not well
defined and states with different numbers of atoms have different energies.  
A self-consistent version of the Bogoliubov theory, which
explicitly takes into account that there is a well defined number of atoms
in a trap, was put forward in Ref.~\cite{gardiner} and independently by
Castin and Dum \cite{castin}. In the present section we briefly summarize
the latter results.

   In the formalism of Ref.~\cite{castin} it is assumed that most atoms are
condensed, i.e. they occupy the same condensate wavefunction $\phi_0(\vec x)$.
To the leading order the state of the system is a Fock state
\begin{equation}
|\phi_0:N\rangle~,
\end{equation}
with all $N$ atoms in the condensate wavefunction $\phi_0$. In this Fock
state
\begin{equation}
\langle\phi_0:N|~\hat\psi~|\phi_0:N\rangle=0~,
\end{equation}
as expected for any state with a well defined number of atoms.
Fluctuations around such a perfect condensate are calculated with
the help of an expansion

\begin{equation}
\hat\psi(\vec x)~=~\hat a_0~\phi_0(\vec x)~+~\delta\hat\psi(\vec x)~.
\label{decomposition}
\end{equation}
Here $\hat a_0$ annihilates atoms in the condensate wavefunction $\phi_0$
while the fluctuation operator $\delta\hat\psi$ annihilates in all modes
orthogonal to the condensate. To make sure that the $\phi_0$ in
Eq.(\ref{decomposition}) is indeed the condensate wavefunction it is further
assumed that there is no coherence between the condensed and 
non-condensed parts,

\begin{equation}
\langle~\hat a_0^\dagger~\delta\hat\psi~\rangle~=~0~.
\label{nocoherence}
\end{equation}
This constraint ensures that $\phi_0$ is an eigenstate of the one-body
density operator.

{\bf Perturbative expansion. ---} The theory is build by expansion in powers
of the ``small" $\delta\hat\psi$. This operator is ``small" when density of
depleted (non-condensed) atoms is small as compared to density of the 
condensate,
$\langle\delta\hat\psi^{\dagger}(\vec x)\delta\hat\psi(\vec x)\rangle\ll
N|\phi_0(\vec x)|^2$.

In dimensionless units a Hamiltonian for $N$ interacting atoms in a trapping 
potential $V(\vec x)$ is
\begin{equation}
H=\int d\vec x~
\left[
\frac12 \nabla\hat\psi^{\dagger} \nabla\hat\psi +
V(\vec x) \hat\psi^{\dagger} \hat\psi +
\frac{g}{2} \hat\psi^{\dagger}\hat\psi^{\dagger}\hat\psi\hat\psi
\right]~,
\label{H}
\end{equation}  
where $g$ is proportional to the $s$-wave scattering length.
The Hamiltonian (\ref{H}) is expanded in powers of $\delta\hat\psi$.

  To zero order in $\delta\hat\psi$ the $\phi_0$ satisfies a stationary
GPE

\begin{equation}
-\frac12\nabla^2\phi_0+V(\vec x)\phi_0+Ng_0|\phi_0|^2\phi_0=\mu\phi_0~.
\label{GPE}
\end{equation}
For a $\phi_0$ that solves the GPE the first order term in the expansion of
the Hamiltonian (\ref{H}) in powers of $\delta\hat\psi$ is zero.
For small depletion (small number of atoms depleted from the condensate
wavefunction) and $N\gg 1$ the second order term 
can be {\it approximately} expressed in terms of
a number-conserving operator

\begin{equation}
\hat\Lambda(\vec x)~=~
\frac{1}{N^{1/2}}~a_0^{\dagger}~\delta\hat\psi(\vec x)~.
\label{Lambda}
\end{equation}
As an example of this approximation take a term
\begin{eqnarray}
&&
2g
\delta\hat\psi^{\dagger}
\phi_0^*\hat a_0^{\dagger}
\phi_0\hat a_0
\delta\hat\psi=
\nonumber\\
&&
2gN|\phi_0|^2\frac{ \delta\hat\psi^{\dagger}
                  \hat a_0^{\dagger}
                  \hat a_0
                  \delta\hat\psi }{N}=
\nonumber\\
&&
2gN|\phi_0|^2\hat\Lambda^{\dagger}\hat\Lambda-
2gN|\phi_0|^2\frac{\delta\hat\psi^{\dagger}\delta\hat\psi }{N}\approx
\nonumber\\
&&
2gN|\phi_0|^2\hat\Lambda^{\dagger}\hat\Lambda~.
\end{eqnarray}
The last approximate equality is justified for small depletion.

  After all such approximations the second order term in the expansion
of the Hamiltonian becomes

\begin{eqnarray}
H_2~\approx~\frac12\int d^3\vec x ~(\Lambda^{\dagger},-\Lambda)~{\cal L}~
\left(\begin{array}{c}\Lambda \\ \Lambda^{\dagger}\end{array}\right), 
\label{H2}
\label{L}
\end{eqnarray}
where
\begin{equation}
{\cal L} = \left(
\begin{array}{cc}
H_{GP}+gN\hat Q|\phi_0|^2\hat Q   &   
g N \hat Q\phi_0^2\hat Q^*        \\
-g N \hat Q^*\phi_0^{*2}\hat Q    &   
\left[-H_{GP}-gN \hat Q|\phi_0|^2\hat Q\right]^*
\end{array}
\right)~,
\label{calL}
\end{equation}
and

\begin{equation}
H_{GP}=-\frac12\nabla^2+V(\vec x)+Ng_0|\phi_0|^2~.
\end{equation}
Here $\hat Q$ is a projection operator on the subspace orthogonal
to $\phi_0$,

\begin{equation}
\hat Q~=~1~-~\phi_0~\langle\phi_0|~.
\end{equation}

{\bf Bogoliubov modes. ---} In order to diagonalize ${\cal L}$ one can first 
solve the Bogoliubov-de Gennes equations

\begin{eqnarray}
&&
-\frac12 \nabla^2 U_m +
V(\vec x) U_m +
2g |\phi_0|^2 U_m +
g\phi_0^2 V_m =
\nonumber\\
&&
\mu U_m+\omega_m U_m~,
\label{BdG}\\
&&
-\frac12 \nabla^2 V_m +
V(\vec x) V_m +
2g |\phi_0|^2 V_m +
g(\phi_0^*)^2 U_m =
\nonumber\\
&&
\mu V_m-\omega_m V_m~.\cr
\nonumber
\end{eqnarray}
The eigenmodes of ${\cal L}$ are

\begin{eqnarray}
u_m~=~\hat Q~U_m~~,~~v_m~=~\hat Q~V_m~.
\label{QQ}
\end{eqnarray}
This projection is necessary because the operator ${\cal L}$ in Eq.(\ref{calL})
is different from the differential operator in Eq.(\ref{BdG}).
 $H_2$ is diagonalized by a Bogoliubov transformation

\begin{equation}
\hat{\Lambda}(\vec x)~=~
\sum_m~\hat b_m~u_m(\vec x)~+~\hat b_m^{\dagger}~v_m^*(\vec x)\;.
\label{expb}
\end{equation}
Given the normalization conditions

\begin{equation}
\langle u_m | u_{m'} \rangle ~-~ 
\langle v_m | v_{m'} \rangle ~=~ 
\delta_{mm'}~,
\label{orto}
\end{equation}
$\hat b_m$'s satisfy bosonic commutation relations

\begin{equation}
[\hat b_m,\hat b_{m'}^{\dagger}]=\delta_{mm'}
{\rm~~and~~}
[\hat b_m,\hat b_{m'}]=0~.
\label{comm}
\end{equation}
These commutation relations are valid in the present order of 
the expansion i.e. for small depletion. The $\hat b^\dagger_m$ 
($\hat b_m$) operators create (annihilate) {\it quasiparticles}. 
The diagonalized $H_2$ is a sum over harmonic oscillators

\begin{equation}
H_{\rm B}~=~
\sum_m~\omega_m~\hat b_m^{\dagger}\hat b_m~.
\label{HBog}
\end{equation}
Equations (\ref{comm},\ref{HBog}) define the BT in the {\it quasiparticle}
representation.

The Bogoliubov vacuum is an eigenstate of $H_{\rm B}$ without any
{\it quasiparticles}
\begin{equation}
\hat b_m~|0_b\rangle~=~0~,~ \mbox{for all}~m~.
\label{vacBog}
\end{equation}
In Section~\ref{stanz} we derive the particle representation of
$|0_b\rangle$ 
state and other Bogoliubov eigenstates (i.e. the states that are a result of 
action 
of the creation operators $\hat b^\dagger_m$ on the Bogoliubov vacuum).


\section{ Bogoliubov eigenstates in particle representation }
\label{stanz}


  For small depletion the approximate Hamiltonian is a sum of harmonic
oscillators, see Eq.(\ref{HBog}). The eigenstate of the Hamiltonian
(\ref{HBog}) without any {\it quasiparticles} is the Bogoliubov vacuum, 
compare Eq.(\ref{vacBog}). The formal solution for $|0_b\rangle$ in the
{\it quasiparticle} representation is not suitable for any analysis of the
structure of the eigenstate like e.g. comparison with exact diagonalization 
for model systems. In the following we give an expression for the Bogoliubov 
eigenstates in the particle representation. It allows one to investigate how 
atoms are depleted from the condensate wavefunction when strength of 
interaction between particles increases. 


{\bf $\hat b$'s in terms of $a$'s. ---} Using the expansion in Eq.(\ref{expb})
and the orthogonality relations in Eq.(\ref{orto}) we can express $\hat{b}$'s
in terms of $\hat{\Lambda}$,

\begin{equation}
\hat b_m =
\langle u_m | \hat\Lambda \rangle -
\langle v_m | \hat\Lambda^{\dagger} \rangle~.
\nonumber
\end{equation}
With definitions

\begin{eqnarray}
&& \hat u_m ~ \equiv ~ \langle u_m | \delta\hat\psi \rangle~, \nonumber\\
&& \hat v_m^{\dagger} ~ \equiv ~
                       \langle v_m | \delta\hat\psi^{\dagger} \rangle~.
\end{eqnarray}
and Eq.(\ref{Lambda}) we get the nonlinear Bogoliubov transformation

\begin{equation}
\hat b_m=
\frac{1}{N^{1/2}}
\left( \hat a_0^{\dagger}\hat u_m-
       \hat a_0\hat v_m^{\dagger}  \right)~.
\label{trans}
\end{equation}
$\hat a_0$ annihilates particles in the condensate, while $\hat u_m$ and
$\hat v_m$ annihilate in the modes $u_m(\vec x)$ and $v_m(\vec x)$ which
by construction are orthogonal to the condensate wavefunction
$\phi_0(\vec x)$. 

    We note that the right hand side of the Bogoliubov transformation 
(\ref{trans}) 
between quasiparticles and particles is nonlinear in particle operators. This 
nonlinearity 
is necessary to conserve the total number of atoms. We will effectively invert 
this 
nonlinear transformation when the total number of particles $N$ is even.


{\bf The finite $N$ Bogoliubov vacuum ${\bf |0_b:N\rangle}$. ---} To write
$|0_b\rangle$ in particle representation we need a particle operator
$\hat d^{\dagger}$ that commutes with all {\it quasiparticle} annihilation
operators $\hat b_m$,

\begin{equation}
\left[~ \hat b_m ~,~ \hat{d}^{\dagger} ~\right] ~=~ 0 ~,~
\mbox{ for all}~m~.
\label{bd}
\end{equation}
It turns out that a two particle operator of the form

\begin{equation}
\hat d~=~ \hat a_0 \hat a_0 ~+~
          \sum_{k,l\neq 0}~Z_{kl}~\hat{\tilde{a}}_k \hat{\tilde{a}}_l
\label{d}
\end{equation}
is the smallest particle number solution of Eqs.(\ref{bd}). The double
summation in the definition of $\hat{d}$ in Eq.(\ref{d}) runs over an
orthonormal basis $\{\tilde{\phi}_1(\vec x),\tilde{\phi}_2(\vec
x),\dots\}$ in the subspace orthogonal to the condensate wavefunction
$\phi_0(\vec x)$. With the vanishing commutator in Eq.(\ref{bd}) and for
an even number of atoms $N$ we can write down Bogoliubov vacuum in the
form

\begin{equation}
|0_b:N\rangle ~\propto~ (\hat d^{\dagger})^{\frac{N}{2}}~|0\rangle~,
\label{BogN}
\end{equation}
which is annihilated by every $\hat b_m$, $\hat b_m|0_b:N\rangle=0$.
This particle representation of $|0_b\rangle$ has a finite and well
defined number of particles $N$. 

  Excited states of the Hamiltonian (\ref{HBog}) are created from
$|0_b:N\rangle$ by application of the {\it quasiparticle} creation operators

\begin{equation}
\hat b_m^{\dagger}=
\frac{1}{N^{1/2}}
\left( \hat a_0 \hat u_m^{\dagger} -
       \hat a_0^{\dagger} \hat v_m   \right)~.
\end{equation}

  Equation (\ref{BogN}) gives particle representation of the Bogoliubov
vacuum for even $N$. For an odd N there is no $N$-particle state annihilated
by $\hat b_m$'s in Eq.~(\ref{trans}). This is easy to check for $N=1$.
The ground state of the Bogoliubov Hamiltonian is not annihilated by
$\hat b_m$'s. This is a direct consequence of the fact that $\hat b_m$'s are 
only approximate bosonic operators: the bosonic commutation relations 
(\ref{comm}) are fulfilled only approximately when depletion is small.
For large $N$ there is no significant difference between $N$ and $N+1$ so 
one may apply the results of an even particle system to an odd one. In the 
following we will illustrate this problem with an exactly solvable example.
                                 

{\bf Solution for the matrix ${\bf Z}$. ---} To get an explicit
$|0_b:N\rangle$ we must solve the condition (\ref{bd}) with respect to
$Z_{kl}$

\begin{eqnarray}
&&
0~=~\left[~ \hat b_m ~,~ \hat d^{\dagger} ~\right] ~=~
\label{solbd}\\
&&
\left[\frac{1}{N^{1/2}}
      \left( \hat a_0^{\dagger}\hat u_m-
             \hat a_0\hat v_m^{\dagger}  \right),
      \hat a_0^{\dagger} \hat a_0^{\dagger} +
      \sum_{k,l\neq 0}Z_{kl}\hat{\tilde{a}}_k^{\dagger}
                            \hat{\tilde{a}}_l^{\dagger}
\right]=
\nonumber\\
&&
\frac{\hat a_0^{\dagger}}{N^{1/2}}
\left\{
-2\hat v_m^{\dagger}+
\sum_{k,l\neq 0}Z_{kl}
\left(
[\hat u_m,\hat{\tilde{a}}_k^{\dagger}]\hat{\tilde{a}}_l^{\dagger}+
[\hat u_m,\hat{\tilde{a}}_l^{\dagger}]\hat{\tilde{a}}_k^{\dagger}
\right)
\right\}~.
\nonumber
\end{eqnarray}
It is convenient to define matrices $U$ and $V$,

\begin{eqnarray}
&&
[\hat u_m,\hat{\tilde{a}}_k^{\dagger}] ~=~
\langle u_m|\tilde\phi_k \rangle ~\equiv~
U_{mk}~,
\label{Umk}\\
&&
\hat v_m^{\dagger} ~\equiv~
V_{mk}\hat{\tilde{a}}_k^{\dagger}~.
\label{Vmk}
\end{eqnarray}
Using the completeness of the basis $\{\tilde{\phi}_1,\tilde{\phi}_2,\dots\}$ we 
rewrite
the last line of Eq.(\ref{solbd}) as a matrix equation

\begin{equation}
V~=~U~Z~.
\label{V=UZ}
\end{equation}
When $u_m$'s make a complete basis in the subspace orthogonal to $\phi_0$,
then $U_{mk}$ is an invertible matrix, and we get

\begin{equation}
Z~=~U^{-1}~V~.
\end{equation}
The choice of the actual basis $\{\tilde{\phi}_1,\tilde{\phi}_2,\dots\}$ 
is in principle arbitrary.   


{\bf Diagonalization of ${\bf Z}$. ---} We note that the matrix $Z$
introduced in Eq.(\ref{d}) is symmetric 

\begin{equation}
Z~=~Z^T~. 
\end{equation}
Under an additional assumption that $Z$ is real,

\begin{equation}
Z~=~Z^*~,
\end{equation}
the matrix $Z$ can be diagonalized in an orthonormal eigenbasis
$\{\phi_1,\phi_2,\dots\}$ with real eigenvalues $\lambda_k$, and
annihilation operators $\{\hat a_1,\hat a_2,\dots\}$,

\begin{equation}
\hat d~=~ \hat a_0^2 ~+~
          \sum_{k\neq 0}~\lambda_k~ \hat{a}_k^2 ~.
\label{ddiag}
\end{equation}

 The diagonal $\hat d$ gives the finite-$N$ Bogoliubov vacuum in particle 
representation in the form, see Eq.(\ref{BogN}),

\begin{eqnarray}
|0_b:N\rangle & \sim & 
(\hat d^{\dagger})^{\frac{N}{2}}|0\rangle~\sim~
\label{5758}\\
\sum_{n_0,n_1,\dots=0}^{N/2}
&&
\delta_{N,2n_0+\dots+2n_M}
\frac{\sqrt{(2n_0)!\dots (2n_M)!}}{n_0!\dots n_M!}
\times
\nonumber\\
&&
\lambda_1^{n_1}\dots\lambda_M^{n_M}
|2n_0,\dots,2n_M\rangle~.
\nonumber
\end{eqnarray}
Here for the sake of convenience we truncate to $M$ Bogoliubov modes. The
ket $|2n_0,2n_1,\dots,2n_M\rangle$ is a Fock state with
$2n_0,2n_1,\dots,2n_M$ particles in the eigenbasis
$\{\phi_0,\phi_1,\dots,\phi_M\}$. When necessary the particle
representation (\ref{5758}) can be translated into an $N$-particle
wavefunction.

  As we can see in Eq.(\ref{5758}), every eigenmode can be
occupied only by an even number of particles. Particles are depleted from
the condensate in pairs ($2n_1+\dots+2n_M$ is even), and every
non-condensate eigenmode $\phi_k$ can be occupied only by an even number
of atoms.


{\bf Single particle density matrix.---} The eigenstates
$\{\phi_0,\phi_1,\dots\}$ of the matrix $Z$ turn out to be also
eigenstates of the single particle density matrix,

\begin{eqnarray} 
&&
\rho^{(1)}(x,y)~\equiv~
\langle 0_b:N|
\hat\psi^{\dagger}(x)\hat\psi(y)
|0_b:N\rangle~=~
\nonumber\\
&&
\sum_{k=0}^{\infty}~
\langle 0_b:N|
\hat a_k^{\dagger}\hat a_k
|0_b:N\rangle~
\phi_k^*(x)\phi_k(y)
\label{rho1}
\end{eqnarray}
Off-diagonal elements vanish thanks to the even occupation numbers in the
state (\ref{5758}). Eq.(\ref{rho1}) provides a familiar interpretation of
the eigenmodes $\{\phi_0,\phi_1,\dots\}$.

  The next Section opens our series of examples.

\section{  Double Well   }\label{2W}

   Let us put $N$ atoms in a double well potential. In the tight binding
approximation, when we restrict only to the Hilbert space spanned by the
ground states in each well, this system is described by a boson Hubbard
model

\begin{equation}
H_{\rm 2W}~=
~-~\Omega~
\left(~\hat c_1^{\dagger}\hat c_2~+~\hat c_2^{\dagger}\hat c_1~\right)~+~
\frac12\sum_{j=1,2}~\hat n_j(\hat n_j -1)~.
\label{Hubbard}
\end{equation}
Here $\hat n_j=\hat c_j^{\dagger}\hat c_j$ for $j=1,2$ is an operator of
the number of atoms in the $j$-th well. The $\hat n_j(\hat n_j -1)$ terms 
describe repulsive interactions between atoms in each well. The first term 
in the Hamiltonian is responsible for tunneling between different wells. 
All units were rescaled so that the only dimensionless parameter is the 
tunneling rate $\Omega$.

\begin{figure}
\centering
\includegraphics*[width=8.6cm]{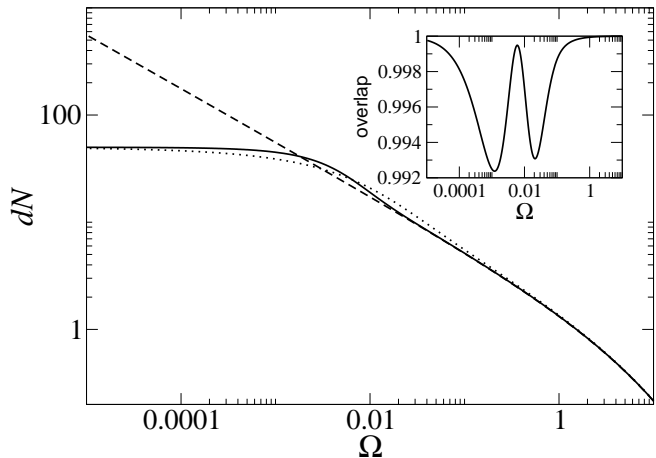}
\caption{  
Number of depleted atoms from a condensate state as a function
of frequency $\Omega$ [see Eq.~({\protect \ref{Hubbard}})].
The solid line corresponds to the exact results, the dashed line
to Eq.~({\protect \ref{dnbog}}), while the dotted line is depletion
calculated directly from the Bogoliubov eigenstate in the particle
representation, Eq.~({\protect \ref{dnz}}). 
The inset shows the overlap of the exact state and the Bogoliubov state 
$|0_b:N\rangle$, Eq.~({\protect \ref{BogN2W}}), as a function of $\Omega$. 
Data for the double well
system with $N=100$ atoms. $\Omega\approx 0.01$ is the crossover
point between the Mott insulator and superfluid. 
}
\label{depl2w}
\end{figure}

\begin{figure}
\centering
\includegraphics*[width=8.6cm]{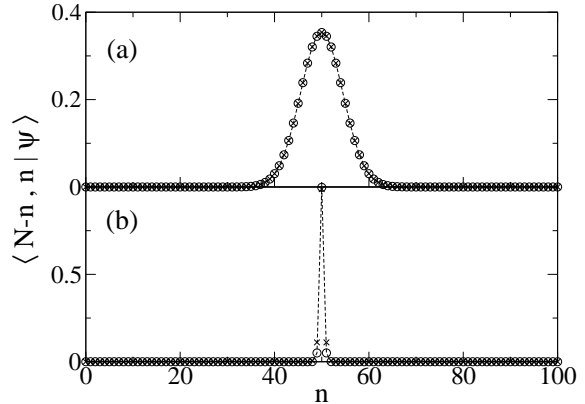}
\caption{ 
Ground state $|\psi\rangle$ of a condensate in a double well potential for 
$N=100$ atoms and for $\Omega=10$ (a), $\Omega=0.001$ (b). Circles (connected 
with a dashed line) correspond to the exact results while crosses are related 
to the Bogoliubov eigenstates in the particle representation. The ground states 
$|\psi\rangle$ are projected on Fock states $|N-n,n\rangle$ where $N-n$ is 
a number of atoms in the left well and $n$ is a number of atoms in the right 
well.     
}
\label{state2w}
\end{figure}

  On the level of the GPE this system is described by a condensate
wavefunction which is a vector of two complex amplitudes,

\begin{equation}
\phi_0~=~\left(~\phi_{0,1}~,~\phi_{0,2}~\right)~,
\end{equation}
one amplitude for each well. Solution of the GPE for the ground state results
in a symmetric state

\begin{equation}
\phi_0~=~\frac{1}{\sqrt{2}}~\left(~1~,~1~\right)~.
\label{cond2W}
\end{equation}
Atoms in this condensate are annihilated by

\begin{equation}
\hat a_0~=~\frac{\hat c_1+\hat c_2}{\sqrt{2}}~.
\end{equation}
The only mode orthogonal to $\phi_0$ is

\begin{equation}
\phi_1~=~\frac{1}{\sqrt{2}}~\left(~1~,~-1~\right)~,
\end{equation}
with an annihilation operator

\begin{equation}
\hat a_1~=~\frac{\hat c_1-\hat c_2}{\sqrt{2}}~.
\end{equation}

  The double well system has been studied theoretically \cite{Javanainen}
and its multiwell generalizations were a subject of recent experiments
\cite{Kasevich}. The ground state properties of this system are well known.
In the regime of $\Omega\gg N$ the ground state is a quasi-coherent state

\begin{equation}
\left(\hat c_1^{\dagger}+\hat c_2^{\dagger}\right)^N|0\rangle~\sim
~\left(\hat a_0^{\dagger}\right)^N|0\rangle~,
\label{quasicoherent}
\end{equation}
where all $N$ atoms are in the same condensate wavefunction
(\ref{cond2W}) and the number of depleted atoms
$N_1=\langle \hat a_1^{\dagger}\hat a_1 \rangle$ is zero.

  In the Mott insulator regime of $N\Omega \ll 1$ the ground state
is a Mott insulator,

\begin{equation}
\left(\hat c_1^{\dagger}\right)^{\frac{N}{2}}
\left(\hat c_2^{\dagger}\right)^{\frac{N}{2}}
|0\rangle~.
\label{Mott}
\end{equation}
This state is very far from the condensate (\ref{quasicoherent}), it has
a huge fraction of depleted atoms
$\frac{\langle \hat c_1^{\dagger}\hat c_1 \rangle}{N}~=~\frac12$.
One half of all atoms are depleted from the condensate wavefunction
(\ref{cond2W}) predicted by the GPE.

{\bf Quasiparticle representation. ---} The Hubbard model
(\ref{Hubbard}) has been studied in the framework of the BT only very
recently \cite{2WBog}. The Bogoliubov-de Gennes equations are

\begin{eqnarray}
&&
-\Omega(u_2-u_1)+\frac{N}{2}(u_1+v_1)=+\omega u_1~,
\nonumber\\
&&
-\Omega(u_1-u_2)+\frac{N}{2}(u_2+v_2)=+\omega u_2~,
\nonumber\\
&&
-\Omega(v_2-v_1)+\frac{N}{2}(v_1+u_1)=-\omega v_1~,
\nonumber\\
&&
-\Omega(v_1-v_2)+\frac{N}{2}(v_2+u_2)=-\omega v_2~,
\label{BogEq2W}
\end{eqnarray}
The index $j=1,2$ in $u_j$ and $v_j$ numbers a well. These equations
give only one Bogoliubov mode orthogonal to the condensate $\phi_0$

\begin{eqnarray}
u~=~\frac{X}{\sqrt{X^2-1}}~\phi_1~,~~v~=~\frac{-1}{\sqrt{X^2-1}}~\phi_1~,
\label{mode}
\end{eqnarray}
where
\begin{equation}
X~=~
\left(1+\frac{4\Omega}{N}\right)+
\sqrt{\left(1+\frac{4\Omega}{N}\right)^2-1}~.
\end{equation}
In the {\it quasiparticle} representation the number of depleted
atoms in the Bogoliubov vacuum state is {\it approximately} given by 
\cite{castin}

\begin{eqnarray}
dN&=&\int dx~\delta\hat\psi^{\dagger}\delta\hat\psi ~\approx~
\int dx~\delta\hat\psi^{\dagger}\hat a_0
\frac{1}{N}\hat a_0^{\dagger}\delta\hat\psi \cr 
&=&
\int dx~\hat\Lambda^{\dagger}\hat\Lambda~
~=~\langle v|v\rangle~=~\frac{1}{X^2-1}~.
\label{dnbog}
\end{eqnarray}
The approximate equality is justified when the depletion is small and $N\gg 1$.
This $dN$ is plotted with a dashed line in Fig.\ref{depl2w}. When 
$\Omega\rightarrow 0$, then the $dN$ diverges like $dN\sim\Omega^{-1/2}$.
As $N$ is finite, the divergence in $dN$ is unphysical. The divergence of 
$dN$ is not present when we calculate depletion directly from the Bogoliubov
state in the particle representation.

\begin{figure}
\centering
\includegraphics*[width=8.6cm]{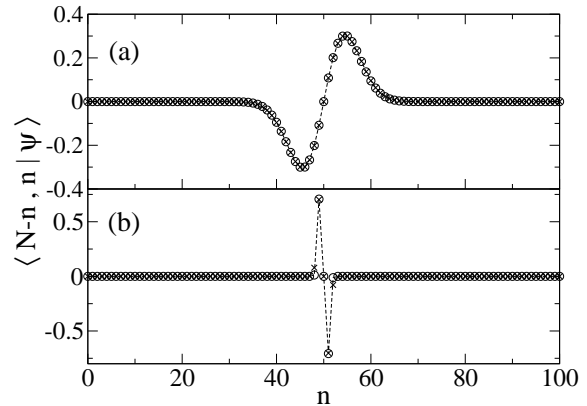}
\caption{ The first excited state $|\psi\rangle$ 
of a condensate in a double well
potential
for $N=100$ atoms and for $\Omega=10$ (a), $\Omega=0.001$ (b).
Circles (connected with a dashed line) correspond to the exact
results while crosses are related to the Bogoliubov eigenstate
with one quasiparticle
$\hat b^\dagger |0_b:N\rangle$.
The states $|\psi\rangle$ are projected on $|N-n,n\rangle$
states where $N-n$ is a number of atoms in the left well and $n$ is a
number
of atoms in the right one.
}
\label{state2wex}
\end{figure}                               

{\bf Particle representation. ---} The operator $\hat b$
for the Bogoliubov mode (\ref{mode}) is

\begin{eqnarray}
\hat b=
\langle u|\Lambda \rangle -
\langle v|\Lambda^{\dagger} \rangle =
\frac{X\hat a_0^{\dagger}\hat a_1+\hat a_1^{\dagger}\hat a_0}
     {\sqrt{N(X^2-1)}}~.
\end{eqnarray}
We look for a two-atom annihilation operator
$\hat d\sim \hat a_0^2+ \lambda_1\hat a_1^2$ such that
$\hat b$ commutes with $\hat d^{\dagger}$, $[\hat b, \hat d^{\dagger}]=0$.
The solution is

\begin{equation}
\hat d\sim
\hat a_0^2-
\frac{\hat a_1^2}{X}~.
\label{d2W}
\end{equation}
This operator defines the Bogoliubov vacuum in the particle representation

\begin{equation}
|0_b:N\rangle\sim\left(\hat d^{\dagger}\right)^{\frac{N}{2}}|0\rangle~.
\label{BogN2W}
\end{equation}
The number of depleted atoms in this state,

\begin{equation}
dN=\langle 0_b:N| \hat a_1^{\dagger}\hat a_1 |0_b:N\rangle~
\label{dnz}
\end{equation}
is plotted with a dotted
line in Fig.\ref{depl2w}. It compares surprisingly well with $dN$ in the
exact ground state (solid line) even in the Mott insulator regime
($N\Omega\ll 1$). In contrast, in the insulator regime
the standard Bogoliubov theory (dashed line) gives an unphysically
divergent $dN$. 

  In Fig.~\ref{state2w} we compare the Bogoliubov eigenstates with the exact 
ones in both weak and strong tunneling regimes. We find surprising agreement 
between both solutions that is present even in the Mott insulator regime. To 
see how the Bogoliubov vacuum smoothly interpolates between the quasi-coherent
state (\ref{quasicoherent}) for large $\Omega$ (large $X$), and 
the Mott insulator (\ref{Mott}) for small $\Omega$ ($X\to 1$) it is
convenient to rewrite Eq.(\ref{d2W}) as a product of annihilation operators
in two in general non-orthogonal modes,

\begin{equation}
d ~\sim~ \left( \hat a_0 + \frac{\hat a_1}{\sqrt{X}} \right)
         \left( \hat a_0 - \frac{\hat a_1}{\sqrt{X}} \right)~.
\end{equation}
With this $d$ the Bogoliubov vacuum becomes a quasi-Fock state

\begin{equation}
|0_b:N\rangle~\sim~
\left( \hat a_0^\dagger + \frac{\hat a_1^\dagger}{\sqrt{X}} 
\right)^{\frac{N}{2}}
\left( \hat a_0^\dagger - \frac{\hat a_1^\dagger}{\sqrt{X}} 
\right)^{\frac{N}{2}}~
|0\rangle~.
\end{equation}
The two modes become the same for large $\Omega$ (large $X$), and they
become orthogonal for $\Omega\to 0$ ($X\to 1$). This quasi-Fock
representation may be useful in any case when one can approximately
truncate the Hilbert space to only one Bogoliubov mode. For the double
well this truncation is exact.

  In Fig.~\ref{state2wex} we compare the first excited state of
the double well system with the Bogoliubov eigenstate with one quasiparticle
$\hat b^{\dagger}|0_b:N\rangle$ for two values of $\Omega$, one
in the Mott insulator and the other in the superfluid regime. Like
for the ground state we find surprisingly good agreement between the
two states even in the Mott regime. 

  The surprising agreement between the Bogoliubov eigenstates and the exact
eigenstates that we found even in the Mott insulator regime is rather an
exception than a rule, as we show for a triple well system.

\section{  Periodic Triple Well }\label{3W}

\begin{figure}
\centering
\includegraphics*[width=8.6cm]{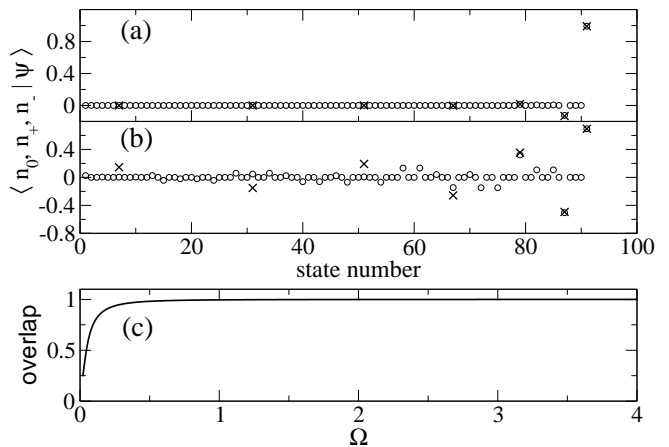}
\caption{ 
Ground state of a condensate in a triple well potential for $N=12$ atoms
and for $\Omega=4$ (a), $\Omega=0.1$ (b). Circles correspond to the exact
results while crosses to the Bogoliubov eigenstates in the particle
representation. The states $|\psi\rangle$ are projected on
$|n_0,n_+,n_-\rangle$ states where $n_0$ is a number of atoms in a
condensate wavefunction Eq.~({\protect \ref{phizero}}) while $n_+$ and
$n_-$ correspond to numbers of atoms in the orthogonal modes $\phi_+$ and
$\phi_-$ Eq.~({\protect \ref{plusminus}}). If the depletion is small (in
(a) $dN\approx 0.03$) the Bogoliubov eigenstates match well the exact
ones. However, for a large depletion (in (b) $dN\approx 1.9$ that is
significant as compared to $N=12$) the exact states become considerably
different from the Bogoliubov prediction that only states $|N-2i,i,i\rangle$
should give nonzero contributions. Panel (c) shows the overlap of the exact 
and Bogoliubov eigenstates as a function of $\Omega$ for $\Omega\in [0.02,4]$.
}
\label{3w}
\end{figure}

\begin{figure}
\centering
\includegraphics*[width=8.6cm]{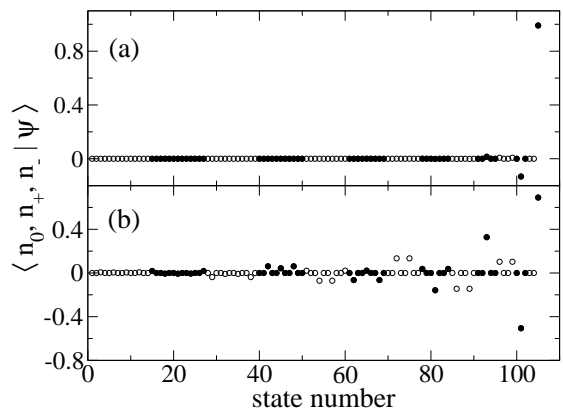}
\caption{ 
Exact ground state of a condensate in a triple well potential for $N=13$
atoms and for $\Omega=4$ (a), $\Omega=0.1$ (b). The states $|\psi\rangle$
are projected on $|n_0,n_+,n_-\rangle$ states where $n_0$ is a number of
atoms in a condensate wavefunction Eq.~({\protect \ref{phizero}}) while
$n_+$ and $n_-$ correspond to numbers of atoms in the orthogonal modes
$\phi_+$ and $\phi_-$ Eq.~({\protect \ref{plusminus}}). Contributions of
states that correspond to even numbers of depleted atoms from the
condensate, i.e. $|N-2i,\ldots,\ldots\rangle$, are marked by full circles.
Compare the structure of the eigenstates with the similar results for
$N=12$ presented in Fig.~{\protect \ref{3w}}.
}
\label{3wodd}
\end{figure}
               
  The Bogoliubov theory applied to the double well potential turns out to
be surprisingly good. In the present example it does not work so well and
we can investigate how the Bogoliubov approximation stops working with
increasing depletion from the condensate. The periodic triple well system
is described by a boson Hubbard Hamiltonian

\begin{equation}
H_{\rm 3W}~=
~-~\Omega~
\sum_{\langle i,j \rangle}
~\hat c_i^{\dagger}\hat c_j~+~
\frac12\sum_{j=1}^3~\hat n_j(\hat n_j -1)~.
\label{Hubbard3w}
\end{equation}
We proceed similarly as for the double well. The ground state of the GPE
is

\begin{equation}
\phi_0=\frac{1}{\sqrt{3}}\left(1,1,1\right)~
\label{phizero}
\end{equation}
and it is associated with an annihilation operator $\hat a_0$. There are 
two modes orthogonal to $\phi_0$

\begin{eqnarray}
&&
\phi_+=
\frac{1}{\sqrt{3}}
\left( 1 , e^{+\frac{2\pi i}{3}} , e^{-\frac{2\pi i}{3}} \right)~,
\nonumber\\
&&
\phi_-=
\frac{1}{\sqrt{3}}
\left( 1 , e^{-\frac{2\pi i}{3}} , e^{+\frac{2\pi i}{3}} \right)~
\label{plusminus}
\end{eqnarray}
with annihilation operators $\hat a_+$ and $\hat a_-$ respectively.
The Bogoliubov-de Gennes equations give two Bogoliubov modes, let us
call them $+$ and $-$, with

\begin{eqnarray}
u_{\pm}=\frac{X\phi_{\pm}}{\sqrt{X^2-1}}~,~~v_{\pm}=\frac{-
\phi_{\pm}}{\sqrt{X^2-1}}~,
\nonumber\\
\end{eqnarray}
where
\begin{equation}
X=
\left(1+\frac{9\Omega}{N}\right)+
\sqrt{ \left(1+\frac{9\Omega}{N}\right)^2 - 1 }~.
\nonumber
\end{equation}
These modes give

\begin{eqnarray}
\hat b_{\pm}=
\frac{X\hat a_0^{\dagger}\hat a_{\pm}+\hat a_0\hat a_{\mp}^{\dagger} }
     {\sqrt{N(X^2-1)}}~.
\end{eqnarray}
The operators $\hat b_{\pm}$ commute with a $\hat d^{\dagger}$ such that

\begin{equation}
\hat d=\hat a_0\hat a_0 - \frac{2}{X} \hat a_+ \hat a_-~,
\end{equation}
which can be easily diagonalized by operators

\begin{eqnarray}
\hat a_1 = \frac{\hat a_+ + \hat a_-}{\sqrt{2}}~,~~
\hat a_2 = \frac{\hat a_+ - \hat a_-}{\sqrt{2}}~,
\end{eqnarray}
with eigenvalues

\begin{equation}
\lambda_1~=~-\lambda_2~=~-\frac{1}{X}~.
\end{equation}

  In the Bogoliubov theory only pairs of atoms are depleted from a
condensate wavefunction. Recently, in exact stochastic calculations
Carusotto and Castin \cite{carusotto} also observed such a pairwise
depletion from a condensate in a harmonic trap. In Fig.~\ref{3w}a we show
the structure of a Bogoliubov vacuum together with the exact ground state
of the triple well system which confirms the predicted pairwise depletion.
However, with decreasing tunneling rate $\Omega$, that makes the
interaction between atoms relatively stronger, the exact ground state
looses the pairwise structure expected in the linear Bogoliubov theory,
see Fig.~\ref{3w}b. This is a clear signature of a breakdown of the
linearized Bogoliubov theory.

  In Fig.~\ref{3wodd} we show an exact ground state for an odd number of
particles. Again like for even $N$ particles are depleted in pairs.  
Although we were not able to find any simple form for the Bogoliubov
ground state with odd $N$, this example suggest that, especially for large $N$,
accurate predictions can be extrapolated from even $N$.

\section{ Bose-Einstein condensate in a harmonic trap }
\label{becharm}

\begin{figure}
\centering
\includegraphics*[width=8.6cm]{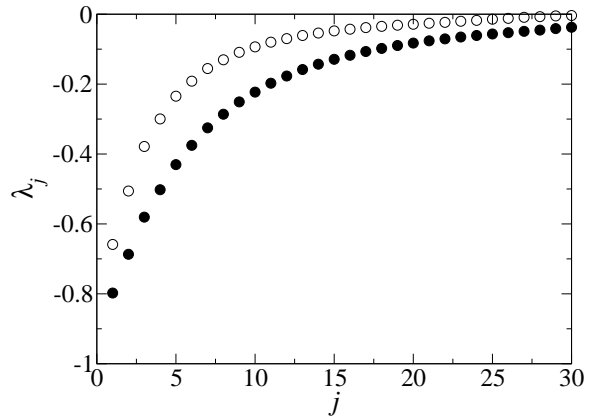}
\caption{ 
Eigenvalues $\lambda_j$ of $Z$ matrix corresponding to the ground state of
a condensate in a 1D harmonic trap for $gN=20$ (open circles) and $gN=50$
(full circles). Note that with increasing $gN$ more and more modes get
significant contributions to a condensate eigenstate, see Eq.~({\protect
\ref{5758}}).
}
\label{becground}
\end{figure}

\begin{figure}
\centering
\includegraphics*[width=8.6cm]{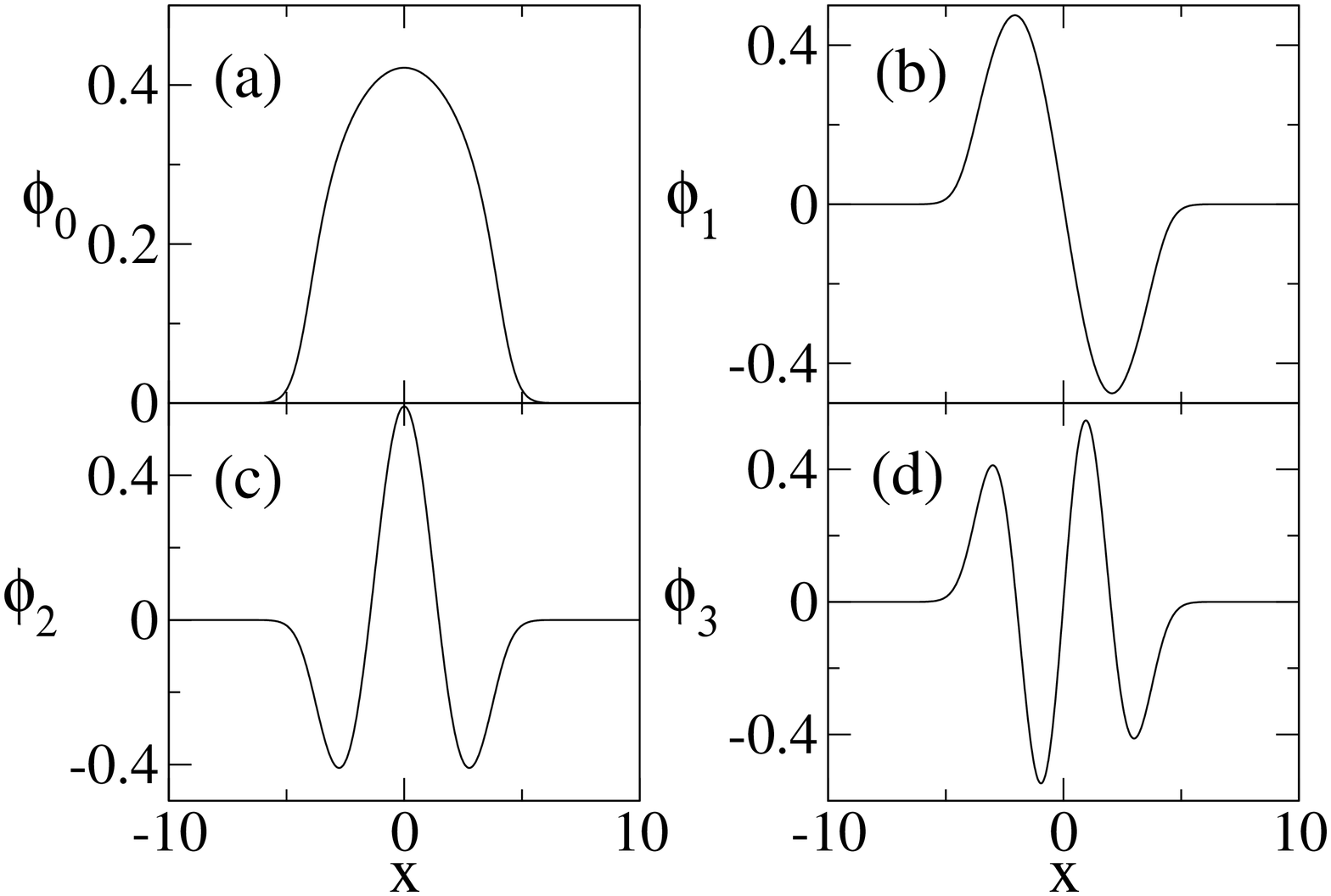}
\caption{
Plots of a condensate mode $\phi_0(x)$ and three the most significant
eigenmodes $\phi_j(x)$ to which the condensate is depleted for the system
in a ground state of a 1D harmonic trap for $gN=50$. Note that
$\phi_j(x)$'s are eigenmodes of $\rho^{(1)}(x,y)$, compare
Eq.(\ref{rho1}).
}
\label{becgroundm}
\end{figure}

\begin{figure}
\centering
\includegraphics*[width=8.6cm]{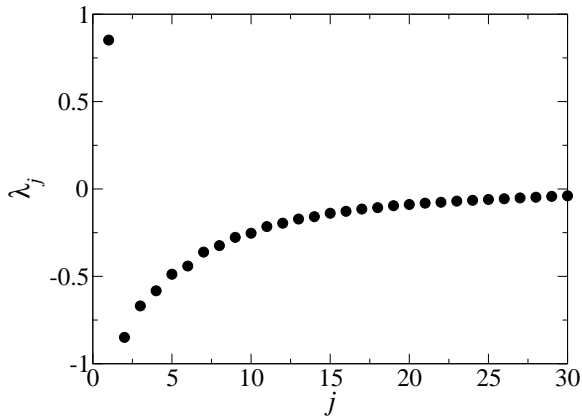}
\caption{ 
Eigenvalues $\lambda_j$ of $Z$ matrix corresponding to a solitonic state
of a condensate in a 1D harmonic trap for $gN=50$. Compare the present
plot with the results for a condensate in a ground state presented in
Fig.~{\protect \ref{becground}}.
}
\label{becsol}
\end{figure}

\begin{figure}
\centering
\includegraphics*[width=8.6cm]{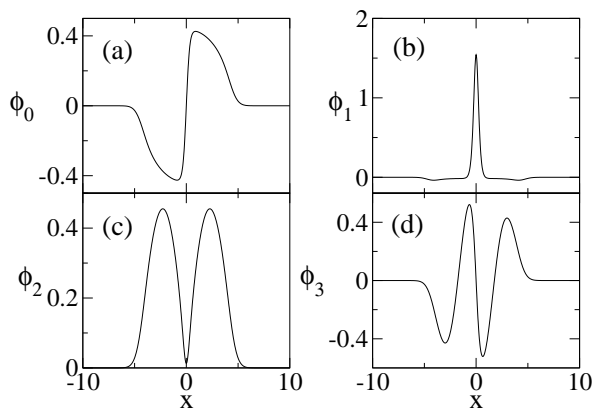}
\caption{ 
Plots of a condensate mode $\phi_0(x)$ and three the most significant
modes $\phi_j(x)$ to which the condensate is depleted for the system in a
solitonic state of a 1D harmonic trap for $gN=50$. Note that $\phi_j(x)$'s
are eigenmodes of $\rho^{(1)}(x,y)$, compare Eq.(\ref{rho1}). 
}
\label{becsolm}
\end{figure}

  In the previous two exactly solvable examples the construction of the
finite $N$ Bogoliubov vacuum was relatively easy because there were only
one or two Bogoliubov modes. In this Section we analyze the finite $N$
Bogoliubov vacuum for ground and solitonic states of a condensate in a 1D
harmonic trap

\begin{equation}
V(x)~=~\frac12 x^2~.
\end{equation}
The solitonic state is the first excited antisymmetric state of the GPE 
\cite{vor-sol-exp}.

  We constructed the Bogoliubov vacuum in particle representation in the
following steps. To begin with we found the relevant stationary state
$\phi_0$ of the GPE in a harmonic potential: the ground state or the
solitonic state. Then we diagonalized the Bogoliubov-de Gennes equations
(\ref{BdG})  for a given background $\phi_0$ and projected the obtained
modes on the subspace orthogonal to the condensate wavefunction like in
Eq.(\ref{QQ}). In this way we obtained the $u_j$ and $v_j$ modes of the
number-conserving Bogoliubov theory. In the next step we constructed
matrices $U$ and $V$ like in Eqs.(\ref{Umk},\ref{Vmk}). In order to construct 
such matrices we had to choose a basis in the subspace orthogonal to the
condensate wavefunction $\phi_0$. To make the matrices finite we had to
truncate the number of Bogoliubov modes to a finite $M$. We got
fast convergence of results with increasing $M$ for a
basis constructed out of the $u_j$ modes,

\begin{eqnarray}
&&
\tilde{\phi}_1~\sim~u_1~,
\nonumber\\
&&
\tilde{\phi}_2~\sim~u_2-\tilde{\phi}_1\langle 
\tilde{\phi}_1|u_2\rangle~,~\dots~.
\end{eqnarray}
Finally, we solved Eq.~(\ref{V=UZ}) with respect to $Z$ and then
diagonalized the obtained $Z$ matrix. The eigenvalues were ordered with
decreasing modulus, $|\lambda_1|>|\lambda_2|>\dots>|\lambda_M|$. As is
clear from the state (\ref{5758}), even a small decrease of $|\lambda_k|$
from one mode to the next results in a substantial drop in the average
number of atoms occupying this mode.

  In Fig.~\ref{becground} we plot the spectrum of the $Z$ matrix,
corresponding to the ground state of the condensate in a harmonic trap for
two different interaction strengths $gN$. With increasing $gN$ more and
more modes have significant contribution to the condensate eigenstate.
Because pairs of atoms are depleted from the condensate, the single
particle eigenmodes $\phi_k$ are allowed to be both even and odd as shown
in Fig.~\ref{becgroundm}. Note that $\phi_j$ are also eigenmodes of the
single particle density matrix, see Eq.(\ref{rho1}).

  For a condensate in a solitonic state, in the spectrum of the $Z$ matrix
(Fig.~\ref{becsol}) the dominant eigenvalue $\lambda_1$ is positive
(contrary to other eigenvalues) and the corresponding anomalous mode is
localized in the soliton notch \cite{dks}, see Fig.~\ref{becsolm}.
The anomalous mode $\phi_1(x)$ is a (nearly) zero mode related to
translational motion of the dark soliton with respect to the
condensate.

\section{ Conclusion }
\label{concl}

   In the present paper we have constructed the Bogoliubov eigenstates in
the particle representation. The particle representation gives much more
intuitive insight into physical processes that are responsible for
depletion of a condensate. Comparison of the derived eigenstates with
exact solutions for model systems reveals that as far as a number of
depleted atoms is small the Bogoliubov eigenstates match well exact
eigenstates. In the Bogoliubov theory the condensate is depleted by pairs of
atoms. However, when strength of the interaction between atoms increases
the linear Bogoliubov theory breaks down and the condensate starts being
depleted by odd numbers of particles.

   We have applied our approach to analyze structure of a realistic
condensate in the ground and solitonic states of a harmonic trap. Having
calculated eigenstates in the particle representation one can in principle
extract any information about a system and simulate results of any
measurement. Examples of such realistic applications will be given in our
future publication \cite{next}.

\section*{Acknowledgments}

Support of KBN under projects 2~P03B~092~23 (J.D.), and 
5~P03B~088~21 (K.S.) is acknowledged.

\end{document}